\documentclass[10pt,aps,prb,reprint,superscriptaddress,nofootinbib,floatfix,longbibliography]{revtex4-2}
\usepackage{amsmath,amsfonts,amsthm,bm} 
\usepackage{amssymb}
\usepackage{amsfonts}
\usepackage{esint}
\usepackage{braket}
\usepackage{comment}
\usepackage{physics}
\usepackage{graphicx}   
\usepackage{dcolumn}    
\usepackage{bm}        
\usepackage{graphicx}  
\usepackage[dvipsnames]{xcolor}
\usepackage[bookmarks=false,linkcolor=Orange,urlcolor=MidnightBlue,colorlinks,citecolor=Maroon]{hyperref}
\usepackage[caption=false]{subfig}

\newcommand{\moire}{moiré }
\newcommand{\Moire}{Moiré }

\usepackage{hyperref}

\begin{document}

\title{Giant Shift Current in Electrically-Tunable Superlattice Bilayer Graphene}
\author{Nabil Atlam}
\email{kamel.n@northeastern.edu}
\affiliation{Northeastern University, Boston, Massachusetts 02115, USA}
\affiliation{Quantum Materials and Sensing Institute, Northeastern University, Burlington, Massachusetts 01803, USA}

\author{Swati Chaudhary}
\email{swatichaudhary@issp.u-tokyo.ac.jp}
\affiliation{The Institute for Solid State Physics, The University of Tokyo, Kashiwa, Chiba 277-8581, Japan}

\author{Arpit Raj}
\affiliation{Northeastern University, Boston, Massachusetts 02115, USA}
\affiliation{Quantum Materials and Sensing Institute, Northeastern University, Burlington, Massachusetts 01803, USA}

\author{Matthew Matzelle}
\affiliation{Northeastern University, Boston, Massachusetts 02115, USA}
\affiliation{Quantum Materials and Sensing Institute, Northeastern University, Burlington, Massachusetts 01803, USA}

\author{Barun Ghosh}
\affiliation{Department of Condensed Matter and Materials Physics S. N. Bose National Center for Basic Sciences, Kolkata-700106, India}

\author{Gregory A. Fiete}
\affiliation{Northeastern University, Boston, Massachusetts 02115, USA}
\affiliation{Quantum Materials and Sensing Institute, Northeastern University, Burlington, Massachusetts 01803, USA}

\author{Arun Bansil}
\email{ar.bansil@northeastern.edu}
\affiliation{Northeastern University, Boston, Massachusetts 02115, USA}
\affiliation{Quantum Materials and Sensing Institute, Northeastern University, Burlington, Massachusetts 01803, USA}

\begin{abstract}
Recent introduction of superlattice potentials has opened new avenues for engineering tunable electronic band structures featuring topologically nontrivial moiré-like bands. Here we consider optoelectronic properties of Bernal-stacked graphene subjected to a superlattice potential either electrostatically or through lattice twisting to show that it exhibits a giant shift current response that is orders of magnitude larger than existing predictions in twisted mulitlayer systems. Effects of gate voltage and the strength and phase of the superlattice potential on the shift current are delineated systematically across various topological regimes.  Our study gives insight into the nature of nonlinear responses of materials and how these responses could be optimized by tuning the superlattice potential. 
\end{abstract}

\maketitle

\section{Introduction}

\Moire materials have emerged as highly tunable platforms for exploring a wide range of intriguing physical phenomena including intrinsic quantum anomalous Hall states \cite{taoValleyCoherentQuantumAnomalous2024, liQuantumAnomalousHall2021}, fractional quantum Hall states \cite{xieFractionalChernInsulators2021, parkObservationFractionallyQuantized2023, zengThermodynamicEvidenceFractional2023, xuObservationIntegerFractional2023, caiSignaturesFractionalQuantum2023}, unconventional superconductivity \cite{caoUnconventionalSuperconductivityMagicangle2018}, correlated insulators \cite{caoCorrelatedInsulatorBehaviour2018}, ferromagnetism \cite{sharpeEmergentFerromagnetismThreequarters2019}, and anomalous Hall crystals \cite{suMoiredrivenTopologicalElectronic2025}. Topologically nontrivial flat bands emerge near charge neutrality in \moire systems and lead to enhanced electronic correlations that drive the formation of various phases with spontaneously broken symmetry \cite{wuTopologicalInsulatorsTwisted2019, bistritzerMoireBandsTwisted2011, crepelAnomalousHallMetal2023, abouelkomsanBandMixingQuantum2024}.
\begin{figure}[ht!]
    \centering
    \includegraphics[width=0.98\columnwidth]{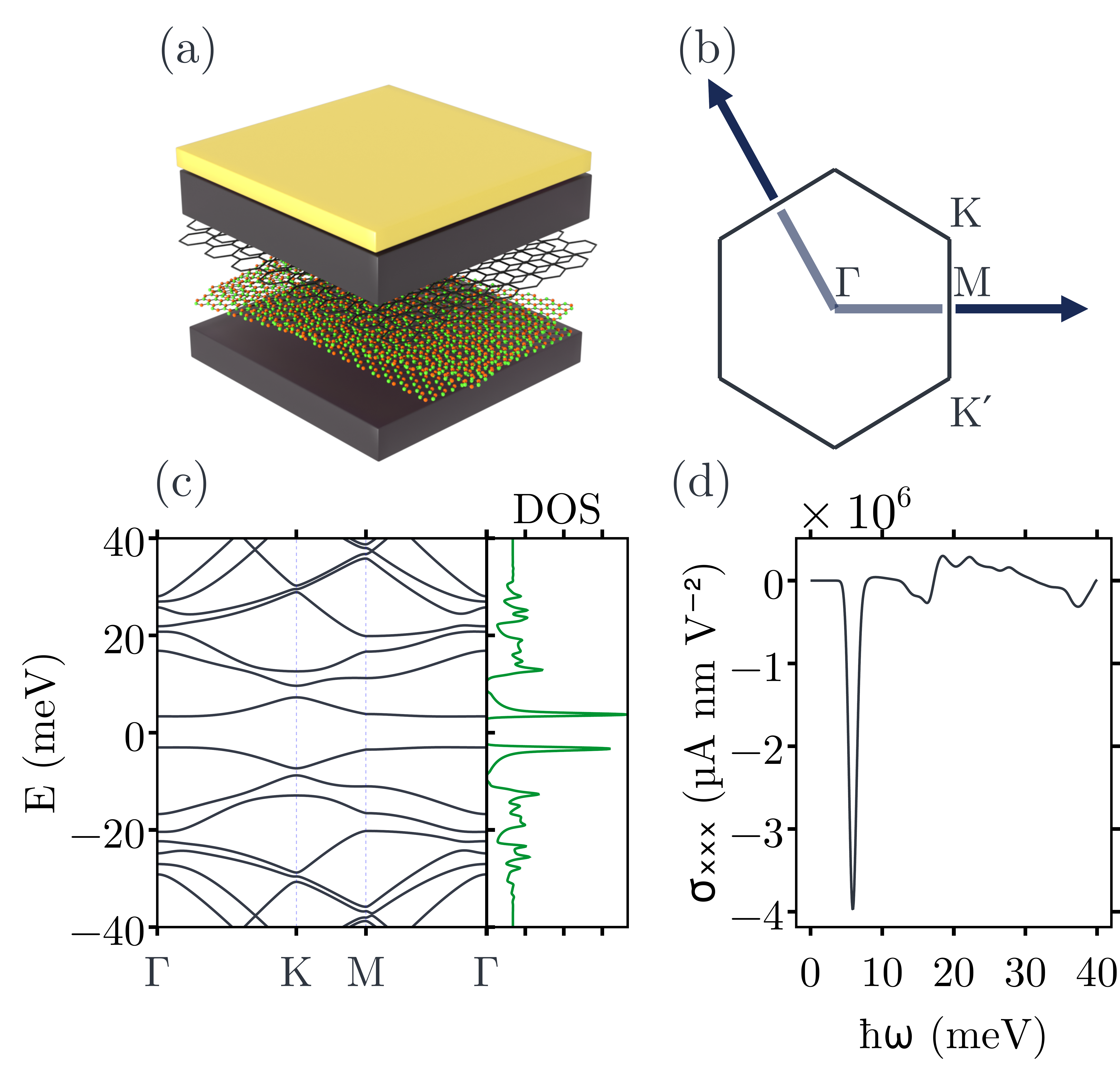}
    \caption{(a) Schematic of a device for generating a \moire potential by placing a layer of twisted hexagonal Boron Nitride (hBN) below a sheet of bilayer graphene. (b) Mini-Brillouin Zone (mBZ) showing four high-symmetry points. (c) Band structure and the associated density of states using a representative set of model parameters ($V_{\text{ESL}} = 5 \ \text{meV}$ and $V_0 = -5 \ \text{meV}$). (d) $\sigma_{xxx}$ component of the shift current conductivity as a function of photon frequency for model parameters in (c)}
    \label{fig:MainFig}
\end{figure}

\Moire materials display a variety of interesting optoelectronic responses \cite{duNonlinearPhysicsMoire2024, duMoirePhotonicsOptoelectronics2023, Topp2021} including higher harmonic generation and the bulk photovoltaic effect (BPVE) in which a non-centrosymmetric material generates a direct photocurrent (DC) \cite{nagaosaNonlinearOpticalResponses2022, nagaosa2017concept, cookDesignPrinciplesShift2017, sipeSecondorderOpticalResponse2000}. In time-reversal symmetric systems, the BPVE arises from the shift-current response to linearly polarized light and the injection-current response to circularly polarized light~\cite{sipeSecondorderOpticalResponse2000,dai2023recent}. The nontrivial quantum geometry of electronic bands~\cite{morimoto2016topological, ahn2022riemannian,morimoto2023geometric,Topp2021,OrensteinJ2021TaSo,ma2023photocurrent} plays a key role here, and the response can surpass the Shockley-Queisser limit \cite{shockleyDetailedBalanceLimit1961}. The bulk photovoltaic effect in \moire materials predominantly occurs in the mid- to far-infrared regions of the electromagnetic spectrum~\cite{Chaudhary2022b,mao2024moir,Kaplan2022TBG,chen2024enhancing,joya2025shift} and the nonlinear optical responses are  more sensitive to crystalline symmetries and the underlying electronic structure than the linear responses~\cite{duNonlinearPhysicsMoire2024}.

A variety of multilayer graphene systems, such as the rhombohedral stacked and twisted multilayer graphene, have been proposed as promising materials for investigating photovoltaic effects driven by quantum geometry~\cite{Chaudhary2022b, Kaplan2022TBG, aroraStraininducedLargeInjection2021, mao2024moir, chen2024enhancing,penaranda2024,Gao2020,postlewaite2024quantum, zheng2023gate,xiong2021atomic,joya2025shift}. Twisted bilayer graphene (TBG) can exhibit a shift current of ~$\sim  10^4 \ \mu\text{A nm V}^{-2}$ in the terahertz (THz) regime, offering the potential to bridge the THz engineering gap \cite{tonouchiCuttingedgeTerahertzTechnology2007}. Shift currents of $10^4$ - $10^5 \ \mu\text{A nm V}^{-2}$ have been predicted in twisted trilayer graphene  \cite{mao2024moir}.  In twisted multilayer systems, the twist angle, number of layers, and the inter-orbital mixing induced by the displacement field can be utilized to optimize the shift-current response~\cite{chen2024enhancing,mao2024moir}.  

An interesting recent method for engineering moiré-like bands involves untwisted bilayer graphene~\cite{wangMoireBandStructure2025, zengGatetunableTopologicalPhases2024, miaoArtificialMoireEngineering2024, tanDesigningTopologyFractionalization2024, Ghorashi2023, seleznevInducingTopologicalFlat2024} where the bilayer graphene sheet is placed in proximity of a periodic electrostatic potential, which allows control over both the symmetry and the geometric parameters of the superlattice potential.  This can be achieved by using a twisted boron nitride substrate \cite{wangMoireBandStructure2025} or by etching a pattern of holes in dielectrics \cite{dubeyTunableSuperlatticeGraphene2013, forsytheBandStructureEngineering2018a,wang2018observation,li2021anisotropic,sun2024signature,shi2019gate,KrixPRB2023}, see Fig.~\ref{fig:MainFig}, opening a new avenue for realizing novel topological state~\cite{miaoArtificialMoireEngineering2024, Ghorashi2023, ault2025optimizing}.

In this study, we discuss shift current in electrostatically engineered AB-stacked (Bernal) bilayer graphene, which is known to exhibit various topological phases~\cite{zheng2023gate,Ghorashi2023}. The tunability of flat bands with superlattice potential is found to offer new opportunities for optimizing the shift current response while circumventing challenges associated with twisted bilayer and multilayer graphene systems.  Our analysis shows that in this way we could realize photo-conductivities even greater than those reported so far in twisted bilayer~\cite{Chaudhary2022b,penarandaPRL2024} and trilayer graphene systems~\cite{postlewaite2024quantum}. 

This paper is organized as follows. Sec.~\ref{sec:BLG} discusses our model system and introduces the associated Hamiltonian. Sec.~\ref{sec:Theory} provides an overview of the relevant theory for the shift current response. Sec.~\ref{sec:results} presents our main results, followed by the conclusions and outlook Sec.~\ref{sec:Discussion}.

\section{AB-Stacked Bilayer Graphene Coupled with a \Moire Potential}
\label{sec:BLG}

We start by considering the following effective Hamiltonian for AB-stacked BLG \cite{BilayerReview}:
\begin{equation}\label{eq:BLGModel}
    \mathcal{H}_{\text{BLG}} = \hbar \nu_F \left(\gamma_z k_x \sigma_x + k_y \sigma_y \right) + \frac{t}{2}\left(\ell_x \sigma_x - \ell_y \sigma_y \right),
\end{equation}
where $\nu_F = 10^6 \ \text{m/s}$ is the Dirac velocity, $t = 0.381 \ \text{eV}$, $\ell_i$, $\sigma_i$, and $\gamma_i$ are the Pauli matrices representing the layer, sublattice, and valley degrees of freedom, respectively. 

Since our focus is on the shift-current conductivity, which does not require breaking of time-reversal symmetry, we project the Hamiltonian of Eq.~\eqref{eq:BLGModel} to the valley sector, where the pseudospin matrix $\gamma_z$ has eigenvalue $+1$. Effects of superlattice potential with both layer-dependent and layer-independent terms are incorporated in the Hamiltonian as follows \cite{zengGatetunableTopologicalPhases2024}:
\begin{align}\label{eq:MoireModel}
    \mathcal{H} &= \mathcal{H}_{\text{BLG}} + V_0 \ell_z  \notag\\
                &  \qquad + \frac{(1 + \alpha) + (1 - \alpha)\ell_z}{2} \, \sum_{n = 0}^2 V_{\text{ESL}} \cos(\mathbf{G_n \cdot r} + \phi),
\end{align}
where $V_0$ is a constant displacement field, $V_{\text{ESL}}$ is the strength of the coupling between the Dirac bands and the modulated displacement field, $\mathbf{G}_n = g\left(\cos{\frac{2\pi n}{3}},\, \sin{\frac{2\pi n}{3}}\right)$ are the scattering vectors of the electrostatic superlattice, $g = \frac{4\pi}{\sqrt{3}L}$, where $L$ is the superlattice period, and $\alpha$ is the ratio between the strengths of the \moire potentials felt by the top and bottom layers. Following Ref.~\cite{zengGatetunableTopologicalPhases2024}, we fix $\alpha=0.9$.

The model described by Eq.~\eqref{eq:MoireModel} has been realized experimentally in Refs.~\cite{forsytheBandStructureEngineering2018a, barconsruizEngineeringHighQuality2022}, where the superlattice term was implemented using a gating technique in which a periodic array of nano-gates was patterned into a dielectric layer. This technique offers a high degree of tunability of the charge density, the displacement field, and the electrostatic superlattice strength. Ref.~\cite{wangMoireBandStructure2025} presents an alternate approach in which the superlattice potential is generated by placing a sheet of twisted hBN) close to graphene. Although the model parameters are less tunable in this case, the generated \moire potential is characterized by $\phi = \frac{\pi}{6}$, which breaks the inversion symmetry of the \moire bands. 
\begin{figure}
    \centering
    \includegraphics[width=1.0\columnwidth]{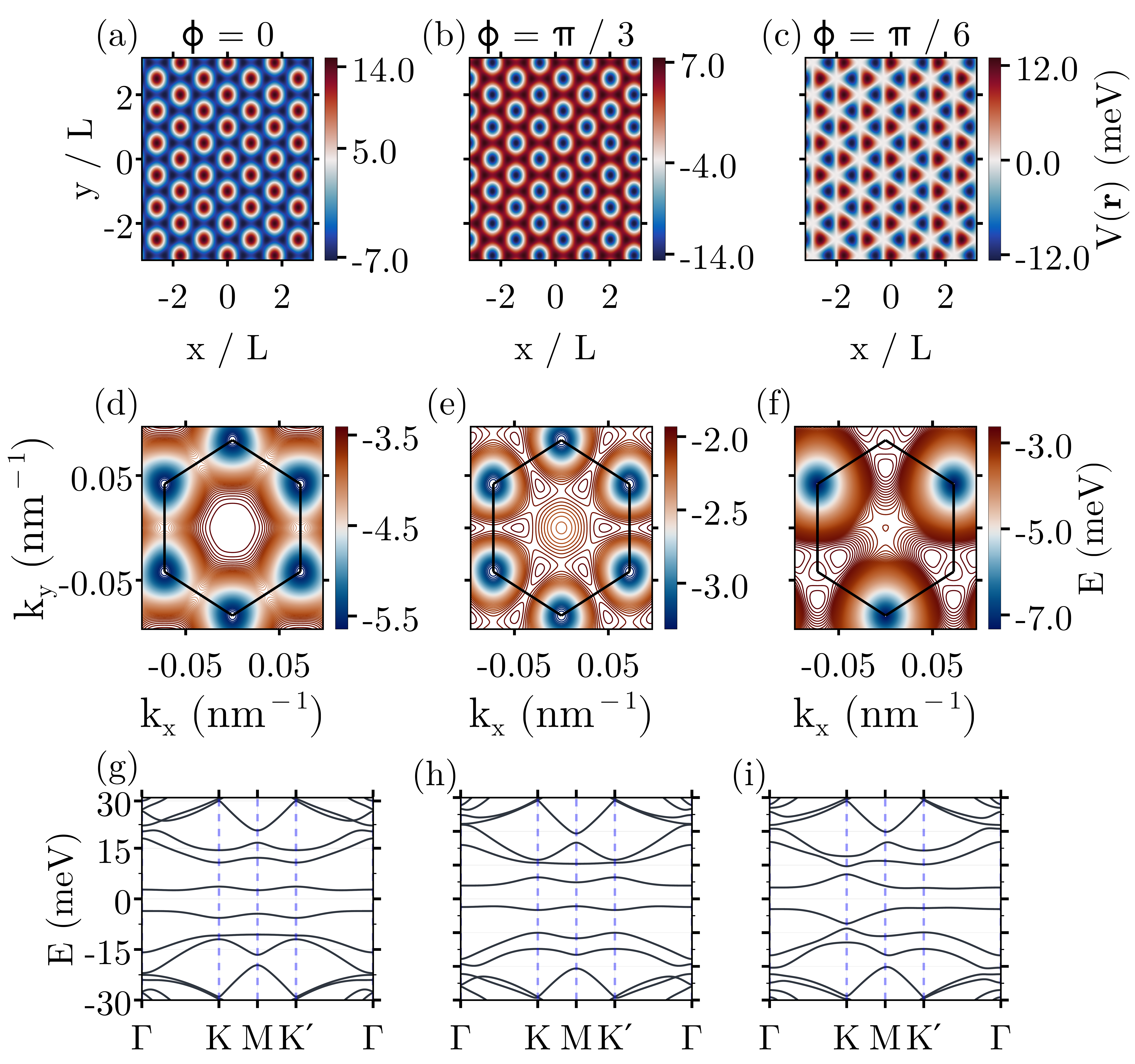}
    \caption{(a-c): Real space profiles of the \moire potential [the sum in Eq.\eqref{eq:MoireModel}] for $V_{\text{ESL}} = 5 \ \text{meV}$ at three different values of $\phi$. For $\phi = 0$, the potential minima form a honeycomb lattice and the potential maxima form a triangular lattice. For $\phi = \frac{\pi}{3}$, the potential minima and maxima form a triangular and a honeycomb lattice respectively. For $\phi = \frac{\pi}{6}$, both the maxima and minima form triangular lattices in real space. (d-f): Contour plots for the top valence band for $V_{\text{ESL}} = 5 \ \text{meV}$, $V_{\text{0}} = -5 \ \text{meV}$, and $L = 50 \ \text{nm}$ for the three $\phi$ values in (a-c). (g-i): Energy bands along high-symmetry lines in the Brillouin zone for three values of $\phi$ as in (a-c), with other parameters fixed to the values in (d-f). In (g) and (h) the band structure is invariant under inversion (compare energy levels at $K$ and $K^{\prime}$ points). In contrast, energy bands for $\phi = \frac{\pi}{6}$ in (i) are not invariant under inversion symmetry.}
    \label{fig:Contours}
\end{figure}

To obtain the energy spectrum, we diagonalize the Hamiltonian of Eq.~\eqref{eq:MoireModel} over the mini-Brillouin zone (mBZ) using a plane wave basis set. Convergence of band structure and shift current calculations was ensured by choosing a high plane-wave energy cutoff of $300-600$ meV. Note that the symmetry of the energy bands depends on the real-space form of the \moire potential. When the strength of the \moire potential is zero, we have full rotational symmetry. Turning on the \moire potential reduces it to six-fold rotational symmetry. Further changes in the symmetry can be induced by varying the phase angle $\phi$. If $\phi$ is an integral multiple of $\frac{\pi}{3}$, then both the \moire potential and the \moire bands possess inversion symmetry, which maps $\mathbf{k} \rightarrow -\mathbf{k}$ and $\mathbf{r} \rightarrow -\mathbf{r}$, see the first two columns of \autoref{fig:Contours}. If $\phi$ is not an integral multiple of $\frac{\pi}{3}$, see the third column of \autoref{fig:Contours}, inversion symmetry is broken. Note that second-order optical responses vanish in materials with inversion symmetry. 

Our one-valley Dirac Hamiltonian possesses additional symmetries. In particular, it is invariant under the anti-unitary $\mathcal{PT}$ symmetry, which conjugates the Hamiltonian by $\ell_x \sigma_x$ together with a complex conjugation. This symmetry is broken by both the displacement field term $V_0 \ell_z$ and the \moire superlattice coupling when $\alpha \neq 1$. \\

\section{An Overview of the Formalism for Nonlinear Optical Conductivity and Shift Current Calculations}\label{sec:Theory}
For light incident normally on a two-dimensional material, the second-order DC photocurrent $J_{a}$ along direction $a$ is given by: $J_{a} = \sum_{b, c}\sigma_{abc}(0; \ \omega, -\omega) E_{b}(\omega)E_{c}(-\omega)$, where $E_a(\omega)$ is the electric field in the $a$ direction at frequency $\omega$ and $\sigma_{abc}$ is the second-order DC current response. The shift current contribution to $\sigma_{abc}$ is defined as \cite{sipeSecondorderOpticalResponse2000, ZoneAveragedShiftCurr}
\begin{align}\label{Eq:SCFormula}
    \sigma_{abc}(0;\omega,-\omega) =& \frac{\pi e^3}{\hbar^2}\sum_{n, m}\int_{\text{BZ}} f^{\text{FD}}_{mn} \left(\mathcal{R}_{mn}^{ab} - \mathcal{R}_{nm}^{ac}\right) r^{c}_{nm}r^{b}_{mn} \notag\\
    &\quad \times \delta(\mathbf{\omega}_{nm} - \omega),
\end{align}
where $n, m$ label the bands with energies $\hbar\omega_n$ and $\hbar\omega_m$, respectively, $\int_{\text{BZ}}=\int\dd[2]k/(2\pi)^2$, $\omega_{nm}=\omega_n-\omega_m$, $f^{\text{FD}}_{nm} = f^{\text{FD}}_{n} - f^{\text{FD}}_{m}$ where $f^{\text{FD}}_n=1/(1+e^{\hbar\omega_n/k_BT})$ is the Fermi-Dirac distribution function, $r^{a}_{mn} = i\mel{u_m}{\partial_a}{u_n}$ are the off-diagonal matrix elements of the position operator, and $\mathcal{R}^{ab}_{mn}(\mathbf{k}) = \mathcal{A}^{a}_{m} - \mathcal{A}^{a}_{n} + i \partial_a \log r_{mn}^{b}$ is the shift vector with $\mathcal{A}^a_m=i\mel{u_m}{\partial_a}{u_m}$ being the abelian Berry connection for band $m$. The shift vector $\mathcal{R}$ is usually interpreted as the shift in the center of mass of the electron wave packet during interband optical transitions~\cite{sipeSecondorderOpticalResponse2000, ahnLowFrequencyDivergenceQuantum2020}. In our numerical calculations of the photocurrent, the temperature is set to zero Kelvin, for simplicity, so the Fermi-Dirac distribution function is replaced by the Heaviside step function. 

Since we are considering photocurrent generation due to linearly-polarized light, Eq.~\eqref{Eq:SCFormula} simplifies to: 
\begin{align}\label{Eq:SCFormula_LP1}
    \sigma_{abb}(0; \omega, -\omega) =& \frac{2 \pi e^3}{\hbar^2}\sum_{n, m}\int_{\text{BZ}} f^{\text{FD}}_{mn} \ \text{Im}\, [\mathcal{R}_{mn}^{ab}]\, |r^{b}_{nm}|^2 \notag\\
    &\quad \times \delta(\mathbf{\omega}_{nm} - \omega),
\end{align}
Following Ref.~\cite{ahn2022riemannian}, Eq.~\eqref{Eq:SCFormula_LP1} can be rewritten as:
\begin{align}\label{Eq:SCFormula_LP2}
    \sigma_{abb}(0; \omega, -\omega) =& \frac{2 \pi e^3}{\hbar^2}\sum_{n, m}\int_{\text{BZ}} f^{\text{FD}}_{mn} \ \text{Im}\,[C_{mn}^{bab}]
     \delta(\mathbf{\omega}_{nm} - \omega),
\end{align}
where $C^{bab}_{mn} = r^{b}_{nm} r^{b}_{mn; a}$ is defined in terms of the position operator and its generalized derivative $r^{b}_{mn; a} = \partial_a r^{b}_{mn} -i r_{mn}^{b}\left(\mathcal{A}^{a}_{m} - \mathcal{A}^{a}_{n}\right) = -i\, \mathcal{R}^{ab}_{mn}r^b_{mn}$. $C^{bab}_{mn}$ is the quantum connection.  To efficiently calculate $C^{bab}_{mn}$ and the generalized derivatives $r^b_{mn; a}$, it is more convenient to employ the sum rule~\cite{cookDesignPrinciplesShift2017, ahnLowFrequencyDivergenceQuantum2020}
\begin{align}\label{Eq:SumRules}
\begin{split}
    r^{a}_{mn; b} &= \frac{i}{\omega_{mn}} \Bigg[ \frac{v^{a}_{mn}\Delta^{b}_{mn} + v^{b}_{mn}\Delta^{a}_{mn}}{\omega_{mn}} \\
    &\qquad - w^{ab}_{mn} + \sum_{p \neq m, n} \left(\frac{v^{a}_{mp}v^{b}_{pn}}{\omega_{pn}} - \frac{v^{b}_{mp}v^{a}_{pn}}{\omega_{mp}} \right) \Bigg],
\end{split}
\end{align}
where $v^{a}_{mn}(\mathbf{k}) = \bra{u_{m}(\mathbf{k})}\partial_a H(\mathbf{k})\ket{u_n(\mathbf{k})}$ are the matrix elements of the velocity operators, $H(\mathbf{k})$ is the Bloch Hamiltonian at wave vector $\mathbf{k}$, $\Delta^a_{mn} = v^a_{mm} - v^a_{nn}$, and $w^{ab}_{mn} = \bra{u_{m}(\mathbf{k})}\partial_a \partial_b H(\mathbf{k})\ket{u_n(\mathbf{k})}$. In the \moire models, where the Hamiltonian is at most linear in momentum, the second-order derivatives of the Bloch Hamiltonian vanish, so $w^{ab}_{mn} = 0$. 

The first term in Eq.~\eqref{Eq:SumRules} represents the direct optical transition between bands $m$ and $n$, while the third term represents virtual optical transitions involving intermediary energy bands. In \moire systems, the virtual processes are particularly enhanced \cite{chen2024enhancing}, leading to an amplification of the photocurrent due to the presence of multiple bands near the Fermi level. 

Spatial symmetries play an important role in second-order optical processes. Inversion symmetry forces all components of the second-order response functions to vanish. In systems with three-fold rotational symmetry $C_3$, the following constraints hold
\begin{align}
    \sigma_{xxx} = -\sigma_{xyy}, \\
    \sigma_{yyy} = -\sigma_{yxx},
\end{align}
and therefore, it is sufficient to consider $\sigma_{xxx}$ and $\sigma_{yyy}$. 

\section{Results and Discussion}
\label{sec:results}
\begin{figure}
    \centering
    \includegraphics[width=1.0\columnwidth]{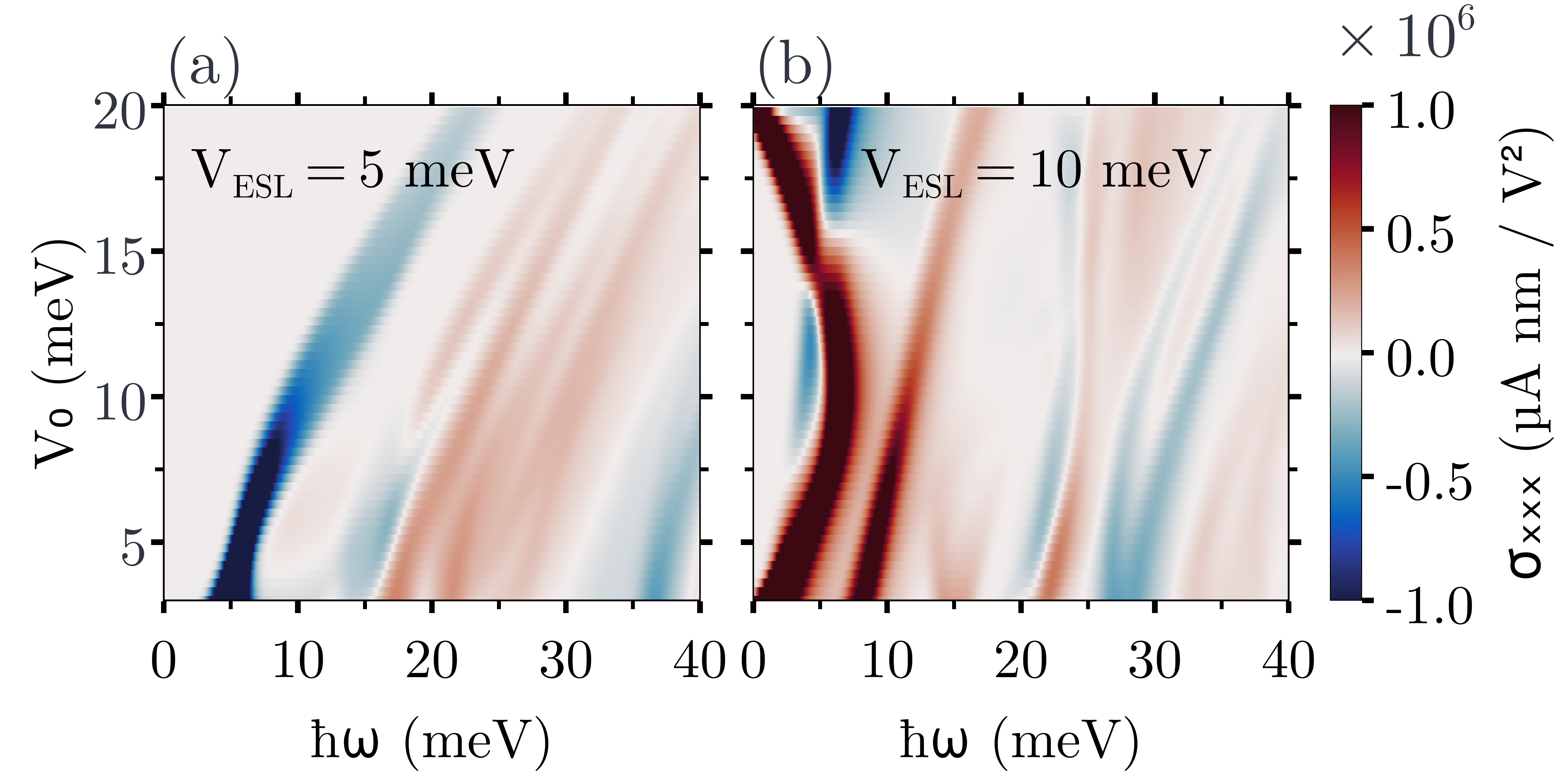}
    \caption{Shift-current photoconductivity dependence on the displacement field for (a) $V_{\text{ESL}} = 5 \ \text{meV}$ and (b) $V_{\text{ESL}} = 10 \ \text{meV}$. Superlattice period and $\phi$ are $50 \ \text{nm}$ and $\frac{\pi}{6}$, respectively. $\sigma_{xxx}$ component of the conductivity tensor is plotted as a function of the incident photon energy $\hbar\omega$ and the displacement field $V_0$. 
    }
    \label{fig:Conxxx_DisplacementField}
\end{figure}

\begin{figure*}[htbp]
    \centering
    \includegraphics[width=\textwidth]{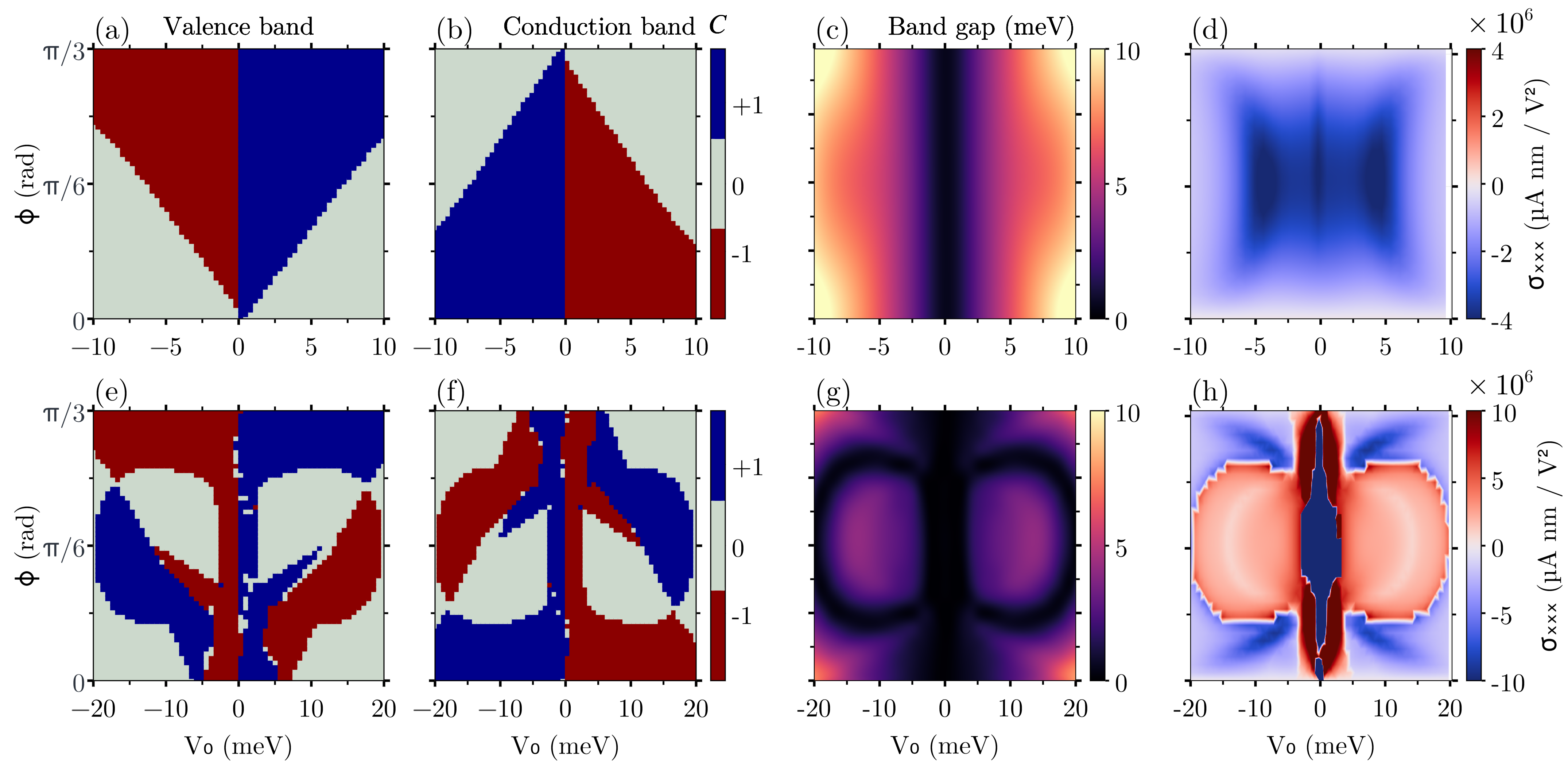}
    \caption{Chern phase diagram, band gap, and the dominant shift current peak as a function of $\phi$ and $V_0$. In all cases, $L = 50 \ \text{nm}$. $V_{\text{ESL}} = 5 \ \text{meV in the top row and} 10 \ \text{meV}$ in the bottom row. (a) and (e): Chern number of the top valence band. (b) and (f): Chern number of the bottom conduction band. (c) and (g): Band gap between the top valence and bottom conduction bands. (d) and (h): Strength of the largest peak in $\sigma_{xxx}$ in units of $\mu \text{A nm V}^{-2}$. Irregularities that are especially noticeable in panels (e) and (f) reflect computational inaccuracies resulting from very small band gaps.}
    \label{fig:ConductivityHeatMap}
\end{figure*}

Results for $\sigma_{xxx}$ and its dependence on representative model parameters are presented in \autoref{fig:MainFig}. Here we focus on the isolated Chern band regime with small \moire potential coupling strengths\footnote{The stacked flat bands regime \cite{Ghorashi2023} will be considered elsewhere.}. The Dirac delta function in Eq.~\eqref{Eq:SCFormula_LP2} was replaced by $\eta \ \pi^{-1} / [(\omega_{nm} -\omega)^2 + \eta^2]$ using the broadening parameter $\eta$ = $0.5 \ \text{meV}$ in all computations\footnote{We take $\eta = 0.5 \ \text{meV}$ following Ref. \cite{penaranda2024}. While the maximum value of the current is sensitive to $\eta$, a larger $\eta$ value broadens the peaks and obscures the distinct spectral features of the shift current }. Notably, due to the spin and valley degeneracies of the \moire bands, value of $\sigma_{xxx}$ is four times the response from the bands in a given spin and valley sector. 

$\sigma_{xxx}$ exhibits several peaks, see \autoref{fig:MainFig}(d), with the dominant peak reaching ~$10^6 \ \mu\text{A nm V}^{-2}$. This large peak results from direct optical transition around $6.5 \ \text{meV}$ between two flat bands near the Fermi energy. [Large density of states (DOS) peaks from these two flat bands can be seen in Fig. 1(c) and correspond to Chern numbers of $-1$ and $+1$.]  Interestingly, the maximum value of shift current in our computations is substantially larger than the corresponding values reported for TBG in the literature of ~$10^4 \ \mu\text{A nm V}^{-2}$ in Ref.~\cite{Chaudhary2022b} and ~$10^5 \ \mu\text{A nm V}^{-2}$ in Ref.~\cite{penarandaPRL2024}. 

Insight into the sensitivity of our results to model parameters is considered in  \autoref{fig:Conxxx_DisplacementField}, which presents shift current as a function of the displacement field $V_{0}$. Value of $\phi ={\pi}/{6}$ used is appropriate for the superlattice potential generated by an hBN film \cite{zhaoUniversalSuperlatticePotential2021}. Other representative values used are: $L = 50 \ \text{nm}$ and $V_{\text{ESL}}= 5 \ \text{meV}$. \autoref{fig:Conxxx_DisplacementField}(a) shows a relatively simple dependence of the photocurrent on $V_{0}$. For small values of $V_{0}$, the current is large and peaked at photon energies of around $5 \ \text{meV}$. With increasing $V_{0}$, the value of current decreases smoothly and the dominant conductivity peak shifts to higher energies, reflecting the monotonic increase in the size of the gap between the two flat bands around the charge neutrality point (CNP) as a function of $V_{0}$. 

\autoref{fig:Conxxx_DisplacementField}(b) considers effects of a higher value of the \moire potential of $V_{\text{ESL}} = 10 \ \text{meV}$ and shows that the dependence of shift current on the displacement field is now more complex. Near $V_0 = 0 $, the shift current has a large value for photon energies below $5 \ \text{meV}$. Increasing $V_0$ initially results in a small decrease in the photocurrent and a shift of the dominant peak towards higher energies, reaching a turning point around $V_0 \sim 12 \ \text{meV}$. Beyond this, the current increases again and the dominant peak shifts to lower photon energies and approaches zero near $V_0 \sim 20 \ \text{meV}$, reflecting the non-monotonic dependence of the band gap on $V_{0}$, see \autoref{fig:ConductivityHeatMap}(g)\footnote{Due to the finite broadening parameter $\eta = 0.5 \ \text{meV}$, a finite zero-frequency response can be observed when the size of the gap is below $\eta$.}. 

\autoref{fig:ConductivityHeatMap} presents effects of varying both the superlattice coupling $\phi$ and the displacement field, $V_0$. Specifically, we focus on the evolution of the two bands closest to zero energy (Fermi energy is placed between these two bands) for $V_0 = -5 \ \text{meV}$ and $\phi = \frac{\pi}{6}$ in the $\phi-V_0$ parameter space, allowing access to different topological regimes. Chern numbers of valence and conduction bands, band gaps, and the strength of the dominant shift-current peak are plotted as a function of $V_0$ and $\phi$ for two different values of the \moire potential strength $V_{\text{ESL}}$ ($5$ and $10 \ \text{meV}$). While the \moire potential generated by a twisted hBN film (Fig. 1) is characterized by $\phi = \frac{\pi}{6}$, other values of $\phi$ could be realized by replacing the hBN film.

\autoref{fig:ConductivityHeatMap}(a, b, e, f) reveal several regions with Chern numbers $+1$, $0$, and $-1$ with boundaries separated by gap closings either between the valence and conduction bands or between two adjacent valence or conduction bands. \autoref{fig:ConductivityHeatMap}(d, h) shows enhancement of the photocurrent near the topological phase boundaries. We also see that the main shift-current peak flips sign across phase boundaries due to gap closings between the valence and conduction bands corresponding to a change in the Chern number of either band between $\pm 1$ and $0$ ~\cite{SignReversal}. This suggests that photocurrent measurements can, in principle, probe topological phases and their boundaries.  

\begin{figure}[!htbp]
    \centering
    \includegraphics[width=1.0\columnwidth]{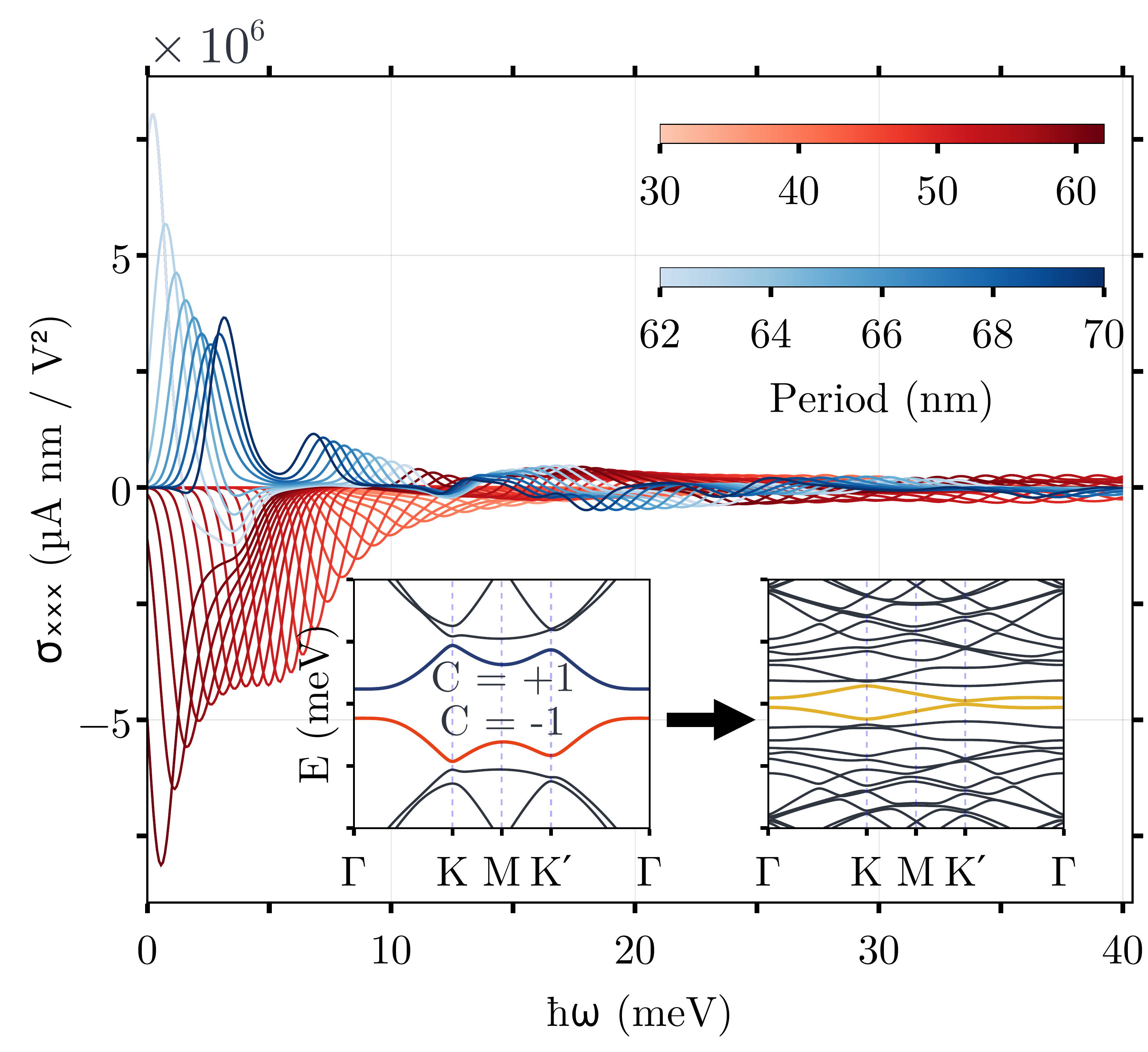}
    \caption{Dependence of $\sigma_{xxx}$ on the \moire period $L$ for fixed values of $V_0=$ $5 \ \text{meV}$ and $\phi$ = ${\pi}/{6}$. \textbf{Inset}: Band structure at two values of the period [$30$ nm (left) and $65 \ \text{nm}$ (right)] showing a topological phase transition, where the Chern numbers of the two bands near the Fermi energy vanish as $L$ increases. }
    \label{fig:Period}
\end{figure}

Finally, \autoref{fig:Period} explores the dependence of the shift current on the \moire period $L$, starting with $L = 30 \ \text{meV}$. Increasing $L$ leads to a compression of the \moire bands and a reduction of the band width as a larger number of bands occupy a smaller energy window around the charge neutrality point. This enhances both direct and virtual optical transitions and leads ultimately to an enhancement of the shift current consistent with our result,  which show a monotonic increase in the shift current (red curves in \autoref{fig:Period})  along with a shift of the dominant peak towards lower photon energies as $L$ increases from $30$ to $60 \ \text{nm}$. At $L \sim 60 \ \text{nm}$, a gap-closing transition occurs at $\rm K^{\prime}$ but not at the $\rm K$ point, which is accompanied with reversal of the sign of the photocurrent and a change in the Chern number of the valence(conduction) band from $-1(+1)$ to $0$. With further increase in $L$, the overall magnitude of the shift current decreases monotonically. Notably, in this regime, quantum geometry of the system becomes sensitive to small changes in parameter values, so that a linear relationship between the photocurrent and the \moire period is not expected \cite{Seleznev2024}.   \\

\section{Conclusions and Outlook}
\label{sec:Discussion}
Following the recent introduction of external superlattice potentials for generating nontrivial topological moiré-like bands, we discuss the DC current response to linearly polarized light (shift current) of a model system based on an AB-stacked graphene bilayer. Effects of gate voltage and the strength and phase of the superlattice potential on the shift current are delineated systematically across various topological regimes. The strength of the shift current in our model is found to be substantially greater than the values reported in the literature for twisted bilayer graphene (TBG). The sign of the main peak in shift current response reverses in going across various topological phase boundaries driven by gap-closings and changes in Chern numbers of the relevant bands. The present materials platform is more tunable in terms of the superlattice potential strength, displacement field, and the carrier charge density of the system and more robust against disorder effects and twist-angle inhomogeneities compared to TBG-based devices. Interesting open areas for further study include exploration of effects of electron-electron interactions not considered in this study and extensions of our model for various multilayer systems and magnetic superlattice potentials for breaking time-reversal symmetry. Our study indicates that band engineering via external superlattice potentials can provide a robust pathway for the development of highly tunable devices for generating large DC photocurrents in the far infrared regime.

\section{Acknowledgments}
Work of N.A., M.M., B.G. and A.B. was supported by the Air Force Office of Scientific Research under award number FA9550-20-1-0322 and benefited from the resources of Northeastern University’s Advanced Scientific Computation Center, the Discovery Cluster, the Quantum Materials and Sensing Institute, and the Massachusetts Technology Collaborative award MTC-22032.  G.A.F. acknowledges funding from the National Science Foundation through DMR-2114825 and additional support from the Alexander von Humboldt Foundation. S.C acknowledges support from JSPS KAKENHI (No. JP23H04865), MEXT, Japan.


\bibliography{bib_files/Nonlinear_Optics,bib_files/Moire, bib_files/intro_moire, bib_files/intro_nonlinear_optics}

\end{document}